
\documentclass[aps,prl,twocolumn]{revtex4-2}

\usepackage[T1]{fontenc}
\usepackage[latin9]{inputenc}
\usepackage{color}
\usepackage{amsmath}
\usepackage{amssymb}
\usepackage{graphicx}
\usepackage[pdftex,unicode=true,pdfusetitle,
 bookmarks=true,bookmarksnumbered=false,bookmarksopen=false,
 breaklinks=false,pdfborder={0 0 1},backref=false,colorlinks=false]
 {hyperref}

\makeatletter

\begin{document}
\title{High temperature superconductivity in the lanthanide hydrides at extreme pressures}
\author{Yao Wei}
\affiliation{King's College London, Theory and Simulation of Condensed Matter
(TSCM), The Strand, London WC2R 2LS, UK}
\author{Francesco Macheda}
\affiliation{King's College London, Theory and Simulation of Condensed Matter
(TSCM), The Strand, London WC2R 2LS, UK}
\author{Zelong Zhao}
\affiliation{King's College London, Theory and Simulation of Condensed Matter
(TSCM), The Strand, London WC2R 2LS, UK}
\author{Terence Tse}
\affiliation{King's College London, Theory and Simulation of Condensed Matter
(TSCM), The Strand, London WC2R 2LS, UK}
\author{Evgeny Plekhanov}
\affiliation{King's College London, Theory and Simulation of Condensed Matter
(TSCM), The Strand, London WC2R 2LS, UK}
\author{Nicola Bonini}
\affiliation{King's College London, Theory and Simulation of Condensed Matter
(TSCM), The Strand, London WC2R 2LS, UK}
\author{Cedric Weber}
\affiliation{King's College London, Theory and Simulation of Condensed Matter
(TSCM), The Strand, London WC2R 2LS, UK}

\date{\today}

\begin{abstract}
Hydrogen-rich superhydrides are promising high-$T_c$ superconductors,
with superconductivity experimentally observed near room
temperature, as shown in recently discovered lanthanide
superhydrides at very high pressures, e.g. LaH$_{10}$ at $170$ GPa and
CeH$_9$ at $150$ GPa. Superconductivity is believed to be closely
related with the high vibrational modes of the bound hydrogen
ions. Here we study the limit of extreme pressures (above 200
GPa) where lanthanide hydrides with large hydrogen content have
been reported. We focus on LaH$_{16}$ and CeH$_{16}$, two prototype
candidates for achieving a large electronic contribution from
hydrogen in the electron-phonon coupling. In this work, we propose
a first-principles calculation platform with the inclusion of
many-body corrections to evaluate the detailed physical properties
of the Ce-H and La-H systems and to understand the structure,
stability and superconductivity of these systems at ultra-high
pressure. We provide a practical approach to further investigate
conventional superconductivity in hydrogen rich superhydrides. We report that density functional theory provides accurate structure and phonon frequencies, 
but many-body corrections lead to an increase of the critical temperature, that is associated with spectral weight transfer of the f-states.
\end{abstract}

\maketitle

\section{Introduction}
In the research of condensed matter~\cite{mcmillan2004condensed}, pressure is a fundamental thermodynamic variable that determines the state of matter, and plays an important role in  the field. The discovery of new materials and thence application to industrial use makes up an important part of modern innovation. As a basic thermodynamic parameter, pressure demonstrates the ability to activate semi-core electrons, empty orbitals, and non-atom-centered quantum orbitals on interstitial sites, changing a given element's chemistry, and thus resulting in a plethora of novel and previously unexpected occurrences, for example the generation of new types of functional materials that deviate from those under atmospheric pressure (101.325 kPa)~\cite{miao2020chemistry}.

Currently, hydrogen-rich materials have garnered much attention in regards to obtaining superconductivity at high temperatures, and has led to much theoretical and experimental work on the search for high-temperature superconductivity of hydrides under high pressure~\cite{wang2018hydrogen}. The chemical pre-compression method, proposed by Neil Ashcroft in 2004, has led to the key realization that hydrogen-rich compounds are a new potential class of high-temperature superconductors~\cite{ashcroft2004hydrogen}.  As investigating the high-temperature superconductivity of metallic hydrogen is highly challenging~\cite{ashcroft1968metallic}, most scholars have instead redirected focus to the synthesis and properties of hydride-rich compounds instead. Recently, the realization of very high temperature superconductivity, at near room temperature, was discovered in hydrogen disulfide~\cite{nakao2019superconductivity,harshman2017compressed} and lanthanum hydrogen~\cite{sun2021high,kostrzewa2020lah,yi2021stability} an important set of milestones.

Hydrogen disulfide was previously believed to undergo dissociation under high pressures, and was not heavily considered as a potential superconductor. However, recent theoretical work has suggested that the dissociation would not occur, but that on the contrary the material becomes superconducting~\cite{li2014metallization}. Inspired by this discovery, Drozdov et al. compressed sulfur hydride in a diamond anvil cell, and has shown that hydrogen sulphide would indeed form a superconductor when compressed with a $T_c$ of up to 200 K~\cite{drozdov2015conventional}.

Armed with this new knowledge, researchers have expanded their search to include lanthanide hydrides under pressure, with notable examples such as La-H and Y-H. The Fermi density and thus superconductivity in La-H systems are determined not only by the $d$ orbital La electrons and the $s$ orbital H electrons, but also by the $f$ orbital electrons of lanthanum~\cite{sun2020second}.

Using DFT, studies have found that the face-centered cubic (FCC) form of LaH$_{10}$ is a good metal, with various bands crossing the Fermi level that forms a high electronic density at the Fermi level~\cite{liu2019microscopic}. 
In contrast, for the YH$_{10}$ system, only the $d$ orbital Y electrons and the $s$ orbital H electrons become major contributors to the Fermi density at high pressures~\cite{heil2019superconductivity}. As external pressure destabilizes the localised La-$4f$ more than the other orbitals (La-$6s$,La-$5d$), the latter populates when pressure is applied, leading to possible novel emergent quantum states associated with the strong electronic correlations.
In particular, it was purported that the La-H system has a unique high-$T_c$ with $f$ electrons at the Fermi level.~\cite{song2020high}.

The effects of electronic localization and hybridization apply for all high-density compounds, and is a hallmark for a wide family of similar materials signified by transition metals, lanthanides, and rare earth elements. Many-body effects give rise to a number of unique and interesting phenomena in similar systems, such as high-temperature superconductivity in cuprates and the metal-insulator transition in vanadates at ambient temperature~\cite{sarma2013many}. The Zaanen-Sawatsky-Allen theory offers a categorization of transition metal periodic solids in terms of their correlation' strengths, and has provided a good knowledge base, though deep understanding of the hydride's characteristics distant from ambient conditions remains hitherto difficult and incomplete~\cite{olalde2011direct}.

Therefore, in order to identify new hydrogen-rich high-$T_c$ superconductors at the lowest possible pressure, the use of quantitative theoretical computations are required. Superconductivity in these potential compounds are mediated by the interaction between highly intense lattice vibrations of hydrogen atoms and the localised electrons. An accurate description of this interaction necessitates fine descriptions of the electronic characteristics, which are infamous for being difficult for such correlated $f$ systems, treating both itinerant and localised electrons on the same footing.

Several theoretical aspects, including the electron-phonon coupling strength $\lambda$, phonon dispersion relations, electron spectral weight, and cross terms between electron-electron, and electron-phonon interactions are corrected with electronic correlations.

Here, we propose a pragmatic first-principles calculation consistent platform that uses many-body corrections for the electronic spectral weight, which then feeds into adjusted estimates of $T_c$. The many-body corrections to phonon dispersion are typically less drastic than the corrections to the electronic spectra, as correlations effects can shift the f- spectral weight over several eV. Furthermore, since their full treatment is beyond reach, the scope of this work is restricted to only correcting the electronic spectra.

In this manuscript, we show that the many-body corrections in $f$ orbital systems could cause significant changes, with spectral weight shifts in the order of one electron volt. We evaluate the precise physical properties of La-H and Ce-H systems, giving special care to the effect of correlations on spectral properties. We find that LaH$_{16}$ and CeH$_{16}$ have stable P6/mmm space group crystal structures at pressures up to $250$ GPa.
Additionally, for the recently discovered Cerium hydride, we predict a comparatively high $T_c$, evaluated using our pre-established hierarchical approach~\cite{plekhanov2021computational}.

\section{Discussion}

The idea that hydrogen-rich compounds could be potential high $T_c$ superconductors dates back to the turn of the millennium, when chemical pre-compression was proposed as a viable way to reduce the metallization pressure of hydrogen in the presence of other elements, leading to observed $T_c$ exceeding 150 K in the LaH$_{16}$ system. This indicates compressed hydrogen-rich compounds as potential room-temperature superconductors.

The genesis of superconductivity in these hydrides is known to originate from electron-phonon interactions. The characteristic phonon frequency, the electron-phonon coupling strength, the density of states at Fermi level, and the Coulomb pseudopotential $\mu^\star$ are the four factors that define $T_c$ according to BCS theory.
Density functional theory, using conventional pseudo-potentials, such as PBE, is widely acknowledged to provide an accurate explanation of lattice dynamics. However, for compounds with weakly hybridized and localized $f$ electrons, such as La and Ce, where many-body corrections are required, DFT is known to have difficulties in dealing with strong electronic correlations.
We apply a density functional theory technique combined with dynamical mean-field theory (DMFT) in this study. The charge and spin local fluctuations, which are important for the local paramagnetic moment of lanthanide elements, are readily corrected by DMFT.
DMFT is used to account for changes in orbital character at the Fermi surface caused by spectral weight transfer related with Hubbard $f$ band splitting. This influences low energy electron-electron scattering processes via phonon momentum transfer, as stated in the Methods section, according to the Allen-Dynes formalism.

\begin{figure*}
\includegraphics[width=\textwidth]{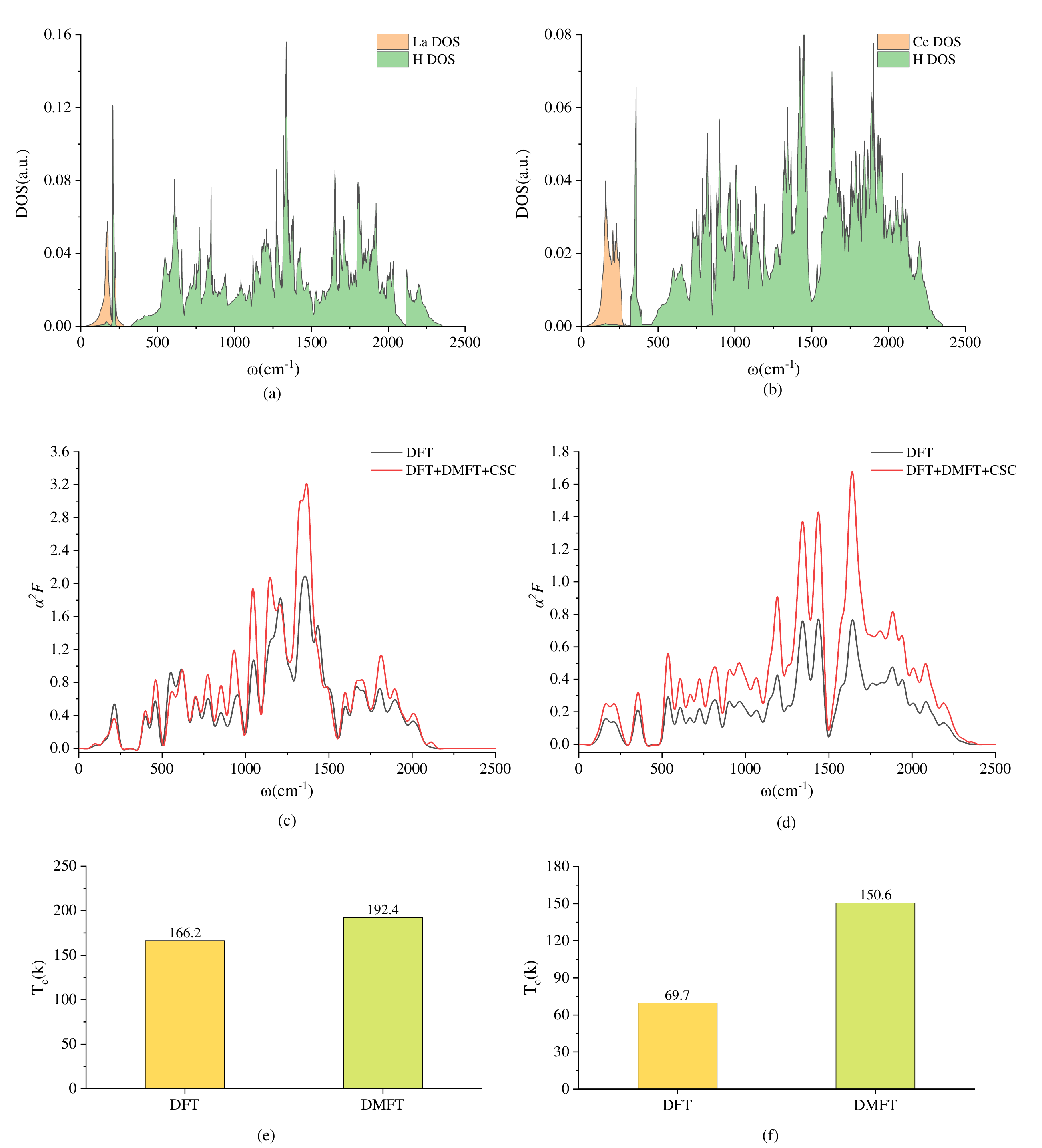}
\caption{\textbf{Many-body corrections to the superconducting temperature}. 
We report (a) phonon density of states, (b)Eliashberg function $\alpha^2F(\omega)$, the spectral weight at the Fermi level is obtained at different levels of approximation: i) DFT PBE (black line),  and ii) with the full charge self-consistent formalism (DFT+DMFT+CSC,  red line).
(c) the superconducting temperature $T_{c}$ obtained by the Allen and Dynes formalism. We obtain a theoretical estimate for LaH$_{16}$ is $T_c=166.2$K by DFT and $T_c=219.6$K by DMFT, also a theoretical estimate for CeH$_{16}$ is $T_c=69.7$K by DFT and $T_c=165.1$K by DMFT
 The full charge self-consistent DMFT provides a large increase of the superconducting temperature, and the physical value of the Hund's coupling for La and Ce is$U=6$eV, $J=0.6$eV. All calculations were performed in the P6/mmm phase of LaH$_{16}$ and CeH$_{16}$ at 250 GPa.}
\label{Ce_LaH phonon}
\end{figure*}

P6/mmm-LaH$_{16}$ and P6/mmm-CeH$_{16}$ shows very little dependence on pressure at a hundred GPa. The phonon DOS of P6/mmm-LaH$_{16}$ and P6/mmm-CeH$_{16}$ computed at $250$ GPa is shown in Fig.~\ref{Ce_LaH phonon}, panels (a) and (b). Then, we investigate the impact of various DMFT electronic charge self-consistency schemes (see Fig.~\ref{Ce_LaH phonon}(c) and Fig.~\ref{Ce_LaH phonon}(d)).We find that correlation effects have a significant impact on $\alpha^2F(\omega)$. In particular, we compare: i) PBE density functional theory, ii)  DFT+DMFT with the full charge self-consistent formalism (DFT+DMFT+CSC). We used the Koster-Slater interaction vertex for the La and Ce correlated manyfold, with typical values for $U=6$ eV, $J=0.6$ eV. Interestingly, the full charge self-consistent DMFT provides a large increase of the superconducting temperature 
{(see panels (e) and (f) in Fig.~\ref{Ce_LaH phonon})}. This confirms that many-body effects have a sizeable contribution to the prediction of the superconducting temperatures in lanthanide hydrides. Note that we use $U=6$ eV, $J=0.6$ eV throughout the rest of the paper.

\begin{figure*}
\includegraphics[width=1\textwidth]{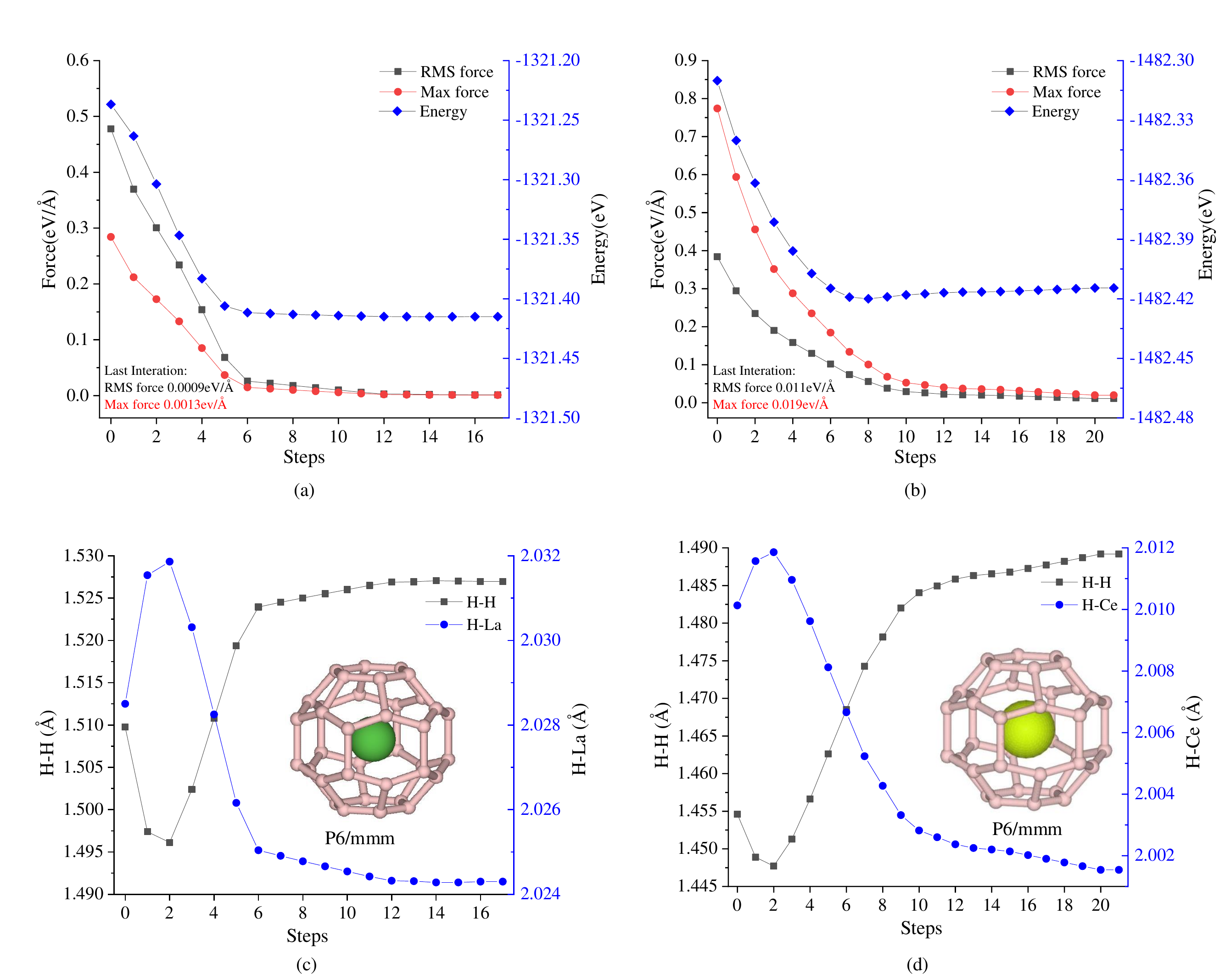}
\caption{\textbf{Structural relaxation of clathrate lanthanides with many-body corrections}. 
We report the structural relaxation of the LaH$_{16}$ (panel a and c) and prototype CeH$_{16}$ (panel b and d) compounds. All calculations are performed at $250$ GPa. The volume density is obtained by the equation of state in DFT+DMFT+CSC that provides very similar results to PBE (not shown). Internal coordinates are relaxed with DFT+DMFT+CSC, building upon the recent implementation of DFT forces for ultra-soft pseudo-potentials. 
We report the forces and total energies obtained during the structural optimization, respectively, in panel (a) (b) for LaH$_{16}$ (CeH$_{16}$). Convergence is obtained within $25$ iterations. The shortest H-H and Ce-H bond lengths increase throughout the structural optimization
(see panel c and d for La and Ce hydrides, respectively).}
\label{DMFT_interation}
\end{figure*}

We focus on the LaH$_{16}$ and CeH$_{16}$ P6/mmm systems at $250$ GPa. Although DMFT readily provides important corrections to the electronic structure, it is worth investigating how DMFT affects structural qualities. Calculating forces with limited atomic displacement is not tractable due to the high computing overhead of doing many-body adjustments.
Recently, we have developed a method for calculation of forces  within DMFT, allowing for ultra-soft and norm-conserving pseudopotentials in the underlying DFT~\cite{plekhanov2021calculating}. This opens up new possibilities for systems with heavy components that are not well-suited to all-electron computations. The structure relaxation at $250$ GPa is seen in Fig.~\ref{DMFT_interation}. Typically, we obtain corrections for the bond lengths of the order of 5$\%$.

The many-body effects tend to slightly shorten the La-H and Ce-H bonds, somehow increasing the La-H and Ce-H covalency. At the same time, this La-H and Ce-H bond length reduction is accompanied by a moderate increase in the H-H distance. This behaviour is opposite to what we have previously found in CeH$_9$, where both Ce-H and H-H bonds were longer within DMFT treatment~\cite{plekhanov2021computational}.

\begin{figure}
\includegraphics[width=1\columnwidth]{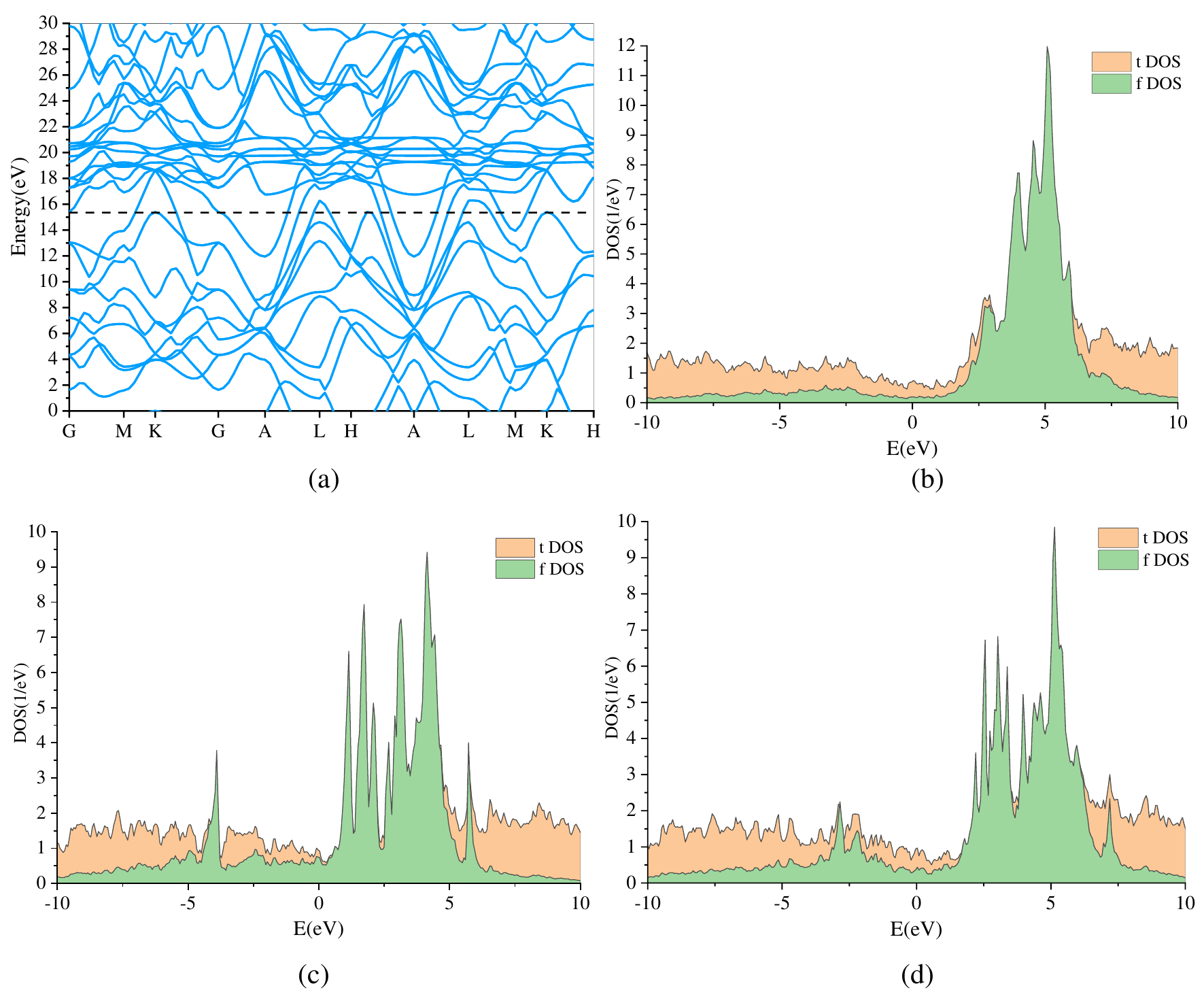}
\caption{\textbf{Spectral weight transfer induced by many-body corrections}.(a) Electronic band structure and (b) density of states obtained by DFT calculations. $t DOS$ and $f DOS$ denote the spectral weight obtained by the imaginary part of respectively the lattice and $f$ impurity Green function, corresponding to the spectral weight traced over all orbitals and traced over the $f$ orbitals, respectively. In (c) and (d) we show the energy-resolved spectral weight, obtained respectively by the one-shot DFT+DMFT and the full charge self-consistent DFT+DMFT+CSC. All calculations were performed in the P6/mmm phase of LaH$_{16}$ at 250 GPa.}
\label{LaH16_dmft}
\end{figure}

\begin{figure}
\includegraphics[width=1\columnwidth]{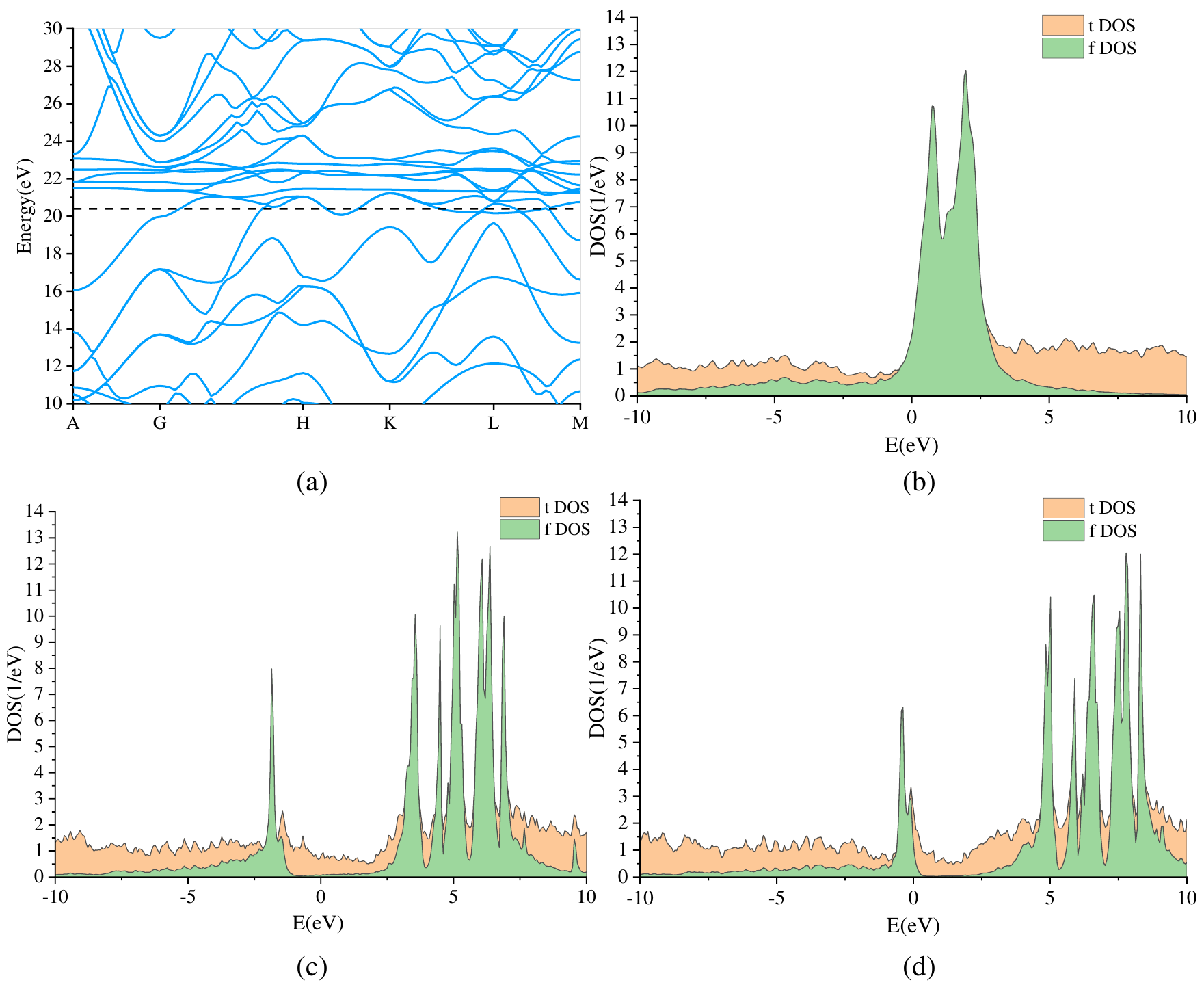}
\caption{\textbf{Spectral weight transfer induced by many-body corrections}.(a) Electronic band structure and (b) density of states obtained by DFT calculations. $t DOS$ and $f DOS$ denote the spectral weight obtained by the imaginary part of respectively the lattice and $f$ impurity Green function, corresponding to the spectral weight traced over all orbitals and traced over the $f$ orbitals, respectively. In (c) and (d) we show the energy-resolved spectral weight, obtained respectively by the one-shot DFT+DMFT and the full charge self-consistent DFT+DMFT+CSC. All calculations were performed in the P6/mmm phase of CeH$_{16}$ at 250 GPa.}
\label{CeH16_dmft}
\end{figure}

The changes highlighted above stem directly from a spectral weight transfer induced by many-body corrections (see Fig. \ref{LaH16_dmft}(a) and \ref{LaH16_dmft}(b)). In DFT, the La system is described by a two band system in absence of long-range magnetic order. We note that DFT is a single Slater determinant approach, and hence cannot capture the role of paramagnetism, with an associated magnetic multiplet (fluctuating magnetic moment). Such effects typically induce a splitting of spectral features into satellites, as observed in Figs. \ref{LaH16_dmft}(c) and \ref{LaH16_dmft}(d), with a resulting large increase of $f$ character at the Fermi level. As sharp La features occur near the Fermi level, we emphasize that a higher level of theory is required to capture correctly the superconducting properties. For instance, in our calculations the one-shot (DFT+DMFT) and full charge self-consistent approach (DFT+DMFT+CSC) induces a small shift of the sharp La feature at the Fermi level, which in turns mitigates the $f$ character increase at the Fermi level. 

The role of $f$ orbitals appears to be very important for the superconducting properties in rare-earth hydrates~\cite{plekhanov2021computational}. That is why in the present paper, in addition to LaH$_{16}$ with formally empty La $f$ shell, we study also Cerium hydrate (CeH$_{16}$) with one $f$ electron in the atomic Ce configuration. We have checked that both systems remain stable at pressure up to at least $250$ GPa.

We report in Fig. \ref{CeH16_dmft} the DFT+DMFT+CSC framework applied to CeH$_{16}$ in the P6/mmm phase at 250 GPa. We attribute the decrease in $T_c$ to a higher $f$ occupation, which shifts the chemical potential away from the $f$ spectral features present near the Fermi level (see Fig.~\ref{CeH16_dmft}(c) and Fig.~\ref{CeH16_dmft}(d) ).

Our findings point to a possible path for increasing $T_c$ in lanthanide hydrides: an increase in $f$ character at the Fermi level in DFT+DMFT, which is associated with a lower degree of La-H and Ce-H covalency and a lower degree of hybridization, which, in turns, is a marker for a higher superconducting temperature in these systems.

\section{Methods}
\begin{figure}
\includegraphics[width=0.9\columnwidth]{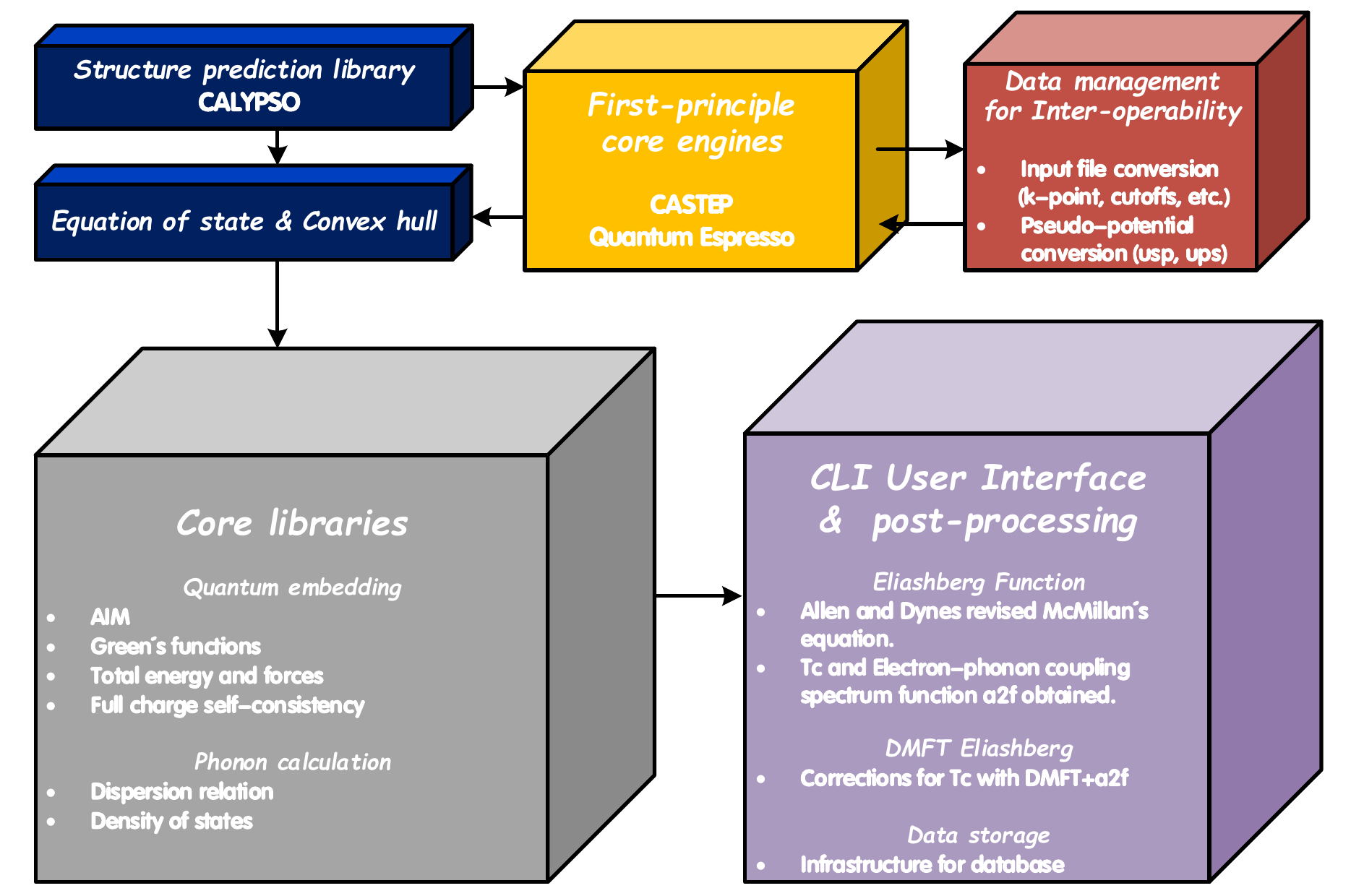}
\caption{\textbf{DFT inter-operability for a consistent many-body platform}. Schematic overview of the main modules of the theoretical platform and its interrelations. Firstly, structures are predicted by Crystal structure AnaLYsis by Particle Swarm Optimization (CALYPSO)), via Gibbs enthalpies for the equation of state and convex hull. The underlying core engines are the CASTEP and Quantum Espresso DFT software. Inter-operability between QE and CASTEP is achieved via format conversion of input files, pseudo-potentials and $\mathbf{k}$-point grids. 
Core libraries are used to provide the many-body corrections, via quantum embedding, which in turns provides corrected forces and energies. In post-processing, the Eliashberg function and superconducting $T_c$ are obtained with the DMFT+a2F approach. Finally, data are archived for future usage in hdf5.}
\label{workflow1pdf}
\end{figure}

Figure \ref{workflow1pdf} depicts our theoretical approach. We present a schematic overview of the theoretical platform's principal elements and their interrelationships. Our method establishes a modular framework for high-pressure material screening. Firstly, Crystal structure AnaLYsis by Particle Swarm Optimization (CALYPSO) provides stoichiometric compositions via Gibbs enthalpies for the equation of state and convex hull~\cite{wang2012calypso}. Interoperability between Quantum Espresso(QE)~\cite{giannozzi2009quantum,clark2005first} is accomplished through input file format conversion, pseudopotentials, and $\mathbf{k}$-point grids. The many-body corrections are provided by core libraries via the DMFT quantum embedding, which results in corrected forces\cite{plekhanov2021calculating} and total free energies\cite{plekhanov2018many,lee2019mott}.

The underlying structure relaxations were carried out using the QE and CASTEP packages in the framework of DFT and using PBE-GGA (Perdew-Burke-Ernzerhof generalized gradient approximation)~\cite{perdew1996generalized,perdew1992atoms}.
Norm conserving pseudopotential were used to describe the core electrons and their effects on valence orbitals\cite{rappe1991erratum}. Valence electron configuration of $5s^2 5p^6 5d^1 6s^2$, $5s^2 5p^6 4f^1 5d^1 6s^2$ (i.e., with explicitly included $f$ electrons) and $1s^1$ was used for the La, Ce and H atoms, respectively. A plane-wave kinetic-energy cut-off of $1000$ eV  and dense Monkhorst-Pack $\mathbf{k}$-points grids with reciprocal space resolution of $12\times 12\times 12$ were employed in the calculation.

Phonon frequencies and superconducting critical temperature were calculated using density-functional perturbation theory as implemented in QE~\cite{baroni2001phonons}. The $\mathbf{k}$-space integration (electrons) was approximated by a summation over a $20\times 20\times 12$ uniform grid in reciprocal space, with the Methfessel-Paxton smearing scheme, using a temperature of $k_B$T = 0.05 eV for self-consistent cycles and relaxations; the same grid ($20\times 20 \times 12$) was used for evaluating DOS and coupling strength. Dynamical matrices and $\lambda$ were calculated on a uniform $5\times 5\times 3$ grid in $\mathbf{q}$-space for P6/mmm-LaH$_{16}$ and P6/mmm-CeH$_{16}$.

In post-processing, the superconducting transition temperature $T_c$ was estimated using the Allen-Dynes modified McMillan equation~\cite{dynes1972mcmillan}:
 
\begin{equation}
T_c = \frac{\omega_{\log}}{1.2} \exp \left [ 
         \frac{-1.04(1+\lambda)}{\lambda - \mu^\star(1+0.62\lambda)}\right ]
\end{equation}
where $\mu^{*}$ is the Coulomb pseudopotential. The electron-phonon coupling strength $\lambda$ and $\omega_{\log}$ were calculated as:
\begin{equation}
\omega_{\log} = \mbox{exp} \left [ \frac{2}{\lambda} \int \frac{d\omega}{\omega}
                                  \alpha^2F(\omega) \log(\omega) \right ],
\end{equation}

\begin{equation}
\lambda = \sum_{\mathbf{q}\nu} \lambda_{\mathbf{q}\nu} = 
2 \int \frac{\alpha^2F(\omega)}{\omega} d\omega.
\end{equation}

In the Allen-Dynes formalism, the Eliashberg function $\alpha^2F(\omega)$ is obtained by summing over all scattering processes at the Fermi level mediated by phonon momentum transfer and reads \cite{allen1975transition}:

\begin{equation}
\alpha^2F(\omega) = N(\epsilon_F) \frac{\sum_{\mathbf{k_1},\mathbf{k_2}}{
                    \left|
                    M_{\mathbf{k_1},\mathbf{k_2}}
                    \right|^2
                    \delta(\omega-\omega_{\mathbf{q}\nu}) \delta(\epsilon_{\mathbf{k_1}}) \delta(\epsilon_{\mathbf{k_2}})}}
                    {\sum_{\mathbf{k_1},\mathbf{k_2}}{
                     \delta(\epsilon_{\mathbf{k_1}}) \delta(\epsilon_{\mathbf{k_2}})}}.
\end{equation}

Here, $N(\epsilon_F)$ is the DOS at Fermi level, $\omega_{\mathbf{q}\nu}$ is the phonon spectrum of a branch $\nu$ at momentum $\mathbf{q}=\mathbf{k_2}-\mathbf{k_1}$, $\epsilon_{\mathbf{k_1}}$ and $\epsilon_{\mathbf{k_2}}$ are electronic band energies referred to the Fermi level, while $M_{\mathbf{k_1},\mathbf{k_2}}$ are the electron-phonon coupling matrix elements. Many-body effects introduce a change of spectral character at the Fermi level, where electronic correlations induce a mass enhancement and introduce a finite lifetime, due to incoherence. In this spirit of the DMFT scissors,
we correct the DFT bands with the renormalised DMFT band picture:
\begin{equation}
\alpha^2F(\omega) = \mathcal{A}_{tot} \frac{\sum_{\mathbf{k_1},\mathbf{k_2}}{
                    \left|
                    M_{\mathbf{k_1},\mathbf{k_2}}
                    \right|^2
                    \delta(\omega-\omega_{\mathbf{q}\nu}) \mathcal{A}(\mathbf{k_1}) \mathcal{A}(\mathbf{k_2})
                    }}
                    {\sum_{\mathbf{k_1},\mathbf{k_2}}{
                     \mathcal{A}(\mathbf{k_1}) \mathcal{A}(\mathbf{k_2})}},
\end{equation}
where $\mathcal{A}_{tot}$ and $\mathcal{A}(\mathbf{k})$ are respectively the total and $\mathbf{k}$-momentum resolved spectral weights at Fermi level. This approach is
denoted as \textit{DMFT+a2F} in the workflow.

Within the DFT+DMFT quantum embedding approach, the DFT Kohn-Sham eigenstates are used in the calculation of the DMFT Green function\cite{weber2021unifying,plekhanov2021computational}. We use atomic projectors to define the Anderson Impurity Model (AIM), which is successively solved within the Hubbard-I approximation. 
A breadth of quantum solvers is readily available in the TRIQS open-source platform\cite{parcollet2015triqs, AICHHORN2016200}. In the full charge self-consistent approach (DFT+DMFT+CSC), the Kohn-Sham potentials are calculated from the DMFT electronic density, obtained using the DMFT occupancies. Upon DMFT convergence, total energies and forces are calculated using the Green function and self-energy. All DFT calculations in this work are carried out using the pseudopotential formalism. 

\section{Conclusion}

We created a methodology for estimating the superconducting temperature in lanthanide hydrides with many-body corrections, and investigated a novel type of stable high-temperature superconducting material, CeH$_{16}$. The DMFT charge self-consistency, which involves many-body adjustments to the local charge density in first-principles calculations, is used to restore a consistent theoretical framework. A change in the spectral weight of the $f$ states causes a rise in predicted superconducting temperature, which influences the spectral character at the Fermi level. We discussed the capabilities for relaxing lanthanide hydrides within the DMFT formalism, built on our recent developments providing DMFT forces for underlying ultra-soft and norm-conserving pseudopotentials, despite the fact that many-body corrections have so far been limited to the electronic contributions to the Eliashberg function.
The latter gives structural insights, and we find that many-body effects on the lattice have negligible impacts at high pressures, since DFT structures and pressures are similar to their DMFT counterparts. Although a comprehensive treatment of phonons at the DMFT level is out of reach for such complicated materials, the latter shows that many-body adjustments to the electronic component are responsible for the strong corrections of $T_c$.
We investigated the aliovalent effect and discovered that compared to iso-structural La hydride and Ce hydride, an increase in $f$ character at the Fermi level leads to an increase in superconducting temperature, which is a compelling observation for future explorations of $f$ systems as high $T_c$ superconductors. Our method is general and provides a modular framework for interoperating common first-principles software, such as the freely accessible CASTEP+DMFT and Quantum Espresso codes, with a tiny numerical footprint and ease of implementation.

\section*{Data Availability}
The codes are available at url \textbf{dmft.ai} under the GPL 3.0 license. 
\quad

\section*{Acknowledgements}
YW are supported from the China Scholarship Council, CW, NB and EP are supported by the grant [EP/R02992X/1] from the UK Engineering and Physical Sciences Research Council (EPSRC). 
This work was performed using resources provided by the ARCHER UK National Supercomputing Service and the Cambridge Service for Data Driven Discovery (CSD3) operated by the University of Cambridge Research Computing Service (www.csd3.cam.ac.uk), provided by Dell EMC and Intel using Tier-2 funding from the Engineering and Physical Sciences Research Council (capital grant EP/P020259/1), and DiRAC funding from the Science and Technology Facilities Council (www.dirac.ac.uk). 

\quad

\section*{Additional Information}
Correspondence should be addressed to Evgeny Plekhanov (evgeny.plekhanov@kcl.ac.uk), Nicola Bonini (nicola.bonini@kcl.ac.uk) and Cedric Weber (cedric.weber@kcl.ac.uk).

\bibliographystyle{apsrev}

\end{document}